\documentclass[journal]{IEEEtran}
\usepackage{ifpdf}
\usepackage{cite}
\usepackage[pdftex]{graphicx}
\usepackage{amsmath}
\usepackage{array}
\usepackage{multirow}
\usepackage{algorithmic}
\usepackage{algorithm}
\usepackage{bbm}
\usepackage{dsfont}
\usepackage{bbold}
\usepackage{multirow}
\usepackage{amssymb}
\usepackage{bm}
\newcommand{\tabincell}[2]{\begin{tabular}{@{}#1@{}}#2\end{tabular}}

\begin{document}
\title{A Spatial-Temporal Graph Neural Network Framework for Automated Software Bug Triaging}
\author{Hongrun~Wu,~\IEEEmembership{Member,~IEEE,}
        Yutao~Ma,~\IEEEmembership{Senior Member,~IEEE,}
        Zhenglong~Xiang,~\IEEEmembership{Member,~IEEE,}
        Chen~Yang,
        and~Keqing~He,~\IEEEmembership{Senior Member,~IEEE}% <-this % stops a space
\thanks{H. Wu is with the College of Physics and Information Engineering, Minnan Normal University, Zhangzhou 363000, China, e-mail: dr.hongrunwu@gmail.com.}% <-this % stops a space
\thanks{Y. Ma and K. He are with the School of Computer Science, Wuhan University, Wuhan 430072, China, e-mails: ytma@whu.edu.cn, hekeqing@whu.edu.cn.}% <-this % stops a space
\thanks{Z. Xiang is with the School of Computer and Software, Nanjing University of Information Science and Technology, Nanjing 210044, China, email: zlxiang@nuist.edu.cn}
\thanks{C. Yang is with the IBO Technology (Shenzhen) Co., Ltd., Shenzhen 212000, China, email: c.yang@ibotech.com.cn} %chen yang
}

\markboth{}%
{Shell \MakeLowercase{\textit{Wu et al.}}: A Spatial-Temporal Graph Neural Network Framework for Automated Software Bug Triaging}
\maketitle

\begin{abstract}
The bug triaging process, an essential process of assigning bug reports to the most appropriate developers, is related closely to the quality and costs of software development. As manual bug assignment is a labor-intensive task, especially for large-scale software projects, many machine-learning-based approaches have been proposed to automatically triage bug reports. Although developer collaboration networks (DCNs) are dynamic and evolving in the real-world, most automated bug triaging approaches focus on static tossing graphs at a single time slice. Also, none of the previous studies consider periodic interactions among developers. To address the problems mentioned above, in this article, we propose a novel spatial-temporal dynamic graph neural network (ST-DGNN) framework, including a joint random walk (JRWalk) mechanism and a graph recurrent convolutional neural network (GRCNN) model. In particular, JRWalk aims to sample local topological structures in a graph with two sampling strategies by considering both node importance and edge importance. GRCNN has three components with the same structure, i.e., hourly-periodic, daily-periodic, and weekly-periodic components, to learn the spatial-temporal features of dynamic DCNs. We evaluated our approach’s effectiveness by comparing it with several state-of-the-art graph representation learning methods in two domain-specific tasks that belong to node classification. In the two tasks, experiments on two real-world, large-scale developer collaboration networks collected from the Eclipse and Mozilla projects indicate that the proposed approach outperforms all the baseline methods.
\end{abstract}


\IEEEpeerreviewmaketitle

\section{Introduction} \label{sec:introduction}

Defect repair (also known as bug fixing) is an everyday activity in software development and maintenance, which plays a crucial role in software quality assurance (SQA). Bug tracking systems, such as Bugzilla\footnote{https://www.bugzilla.org/}, are usually used to track and manage software bugs. Bug triaging, an essential defect assignment process in bug tracking systems, distributes bug reports to the most appropriate developers \cite{anvik2006automating,Anvik2006}. Ideally, one confirmed bug will be efficiently allocated to a proper developer and then quickly fixed by the developer. However, quite a few bug reports are often mistakenly assigned to inappropriate developers in reality, which results in the slow processing speed of bugs in many open-source software projects. The reassignments of a bug report will repeat until a developer eventually fixes the bug. Existing studies show that a large number of bugs were fixed after dozens of reassignments in the Eclipse project, and one reassignment spent about 100 days on average \cite{jeong_improving_2009}.

In recent years, researchers have proposed many automated approaches to reduce human labor costs and improve the bug triaging process’s accuracy and efficiency. In general, these existing studies fall within three main categories. First, the role-based approach \cite{Activity-based,AgrawalATBRT16} labels different developers’ roles, such as reporter, triager, and fixer, and then calculates each developer’s probability of fixing a given bug. Second, the recommendation-based approach \cite{Shokripour2013,jonsson2015automated} uses the machine learning technique to build triage-assisting recommendation models by leveraging bug reports’ structured information, such as title, comment, and description. Third, the hybrid approach combines machine learning and bug tossing graphs (or developer collaboration networks, DCNs) to train developer recommendation models \cite{jeong_improving_2009,Bhattacharya2010Fine,Bhattacharya2012}.

\begin{figure}[hbt!]
\centering
\includegraphics[width=3in]{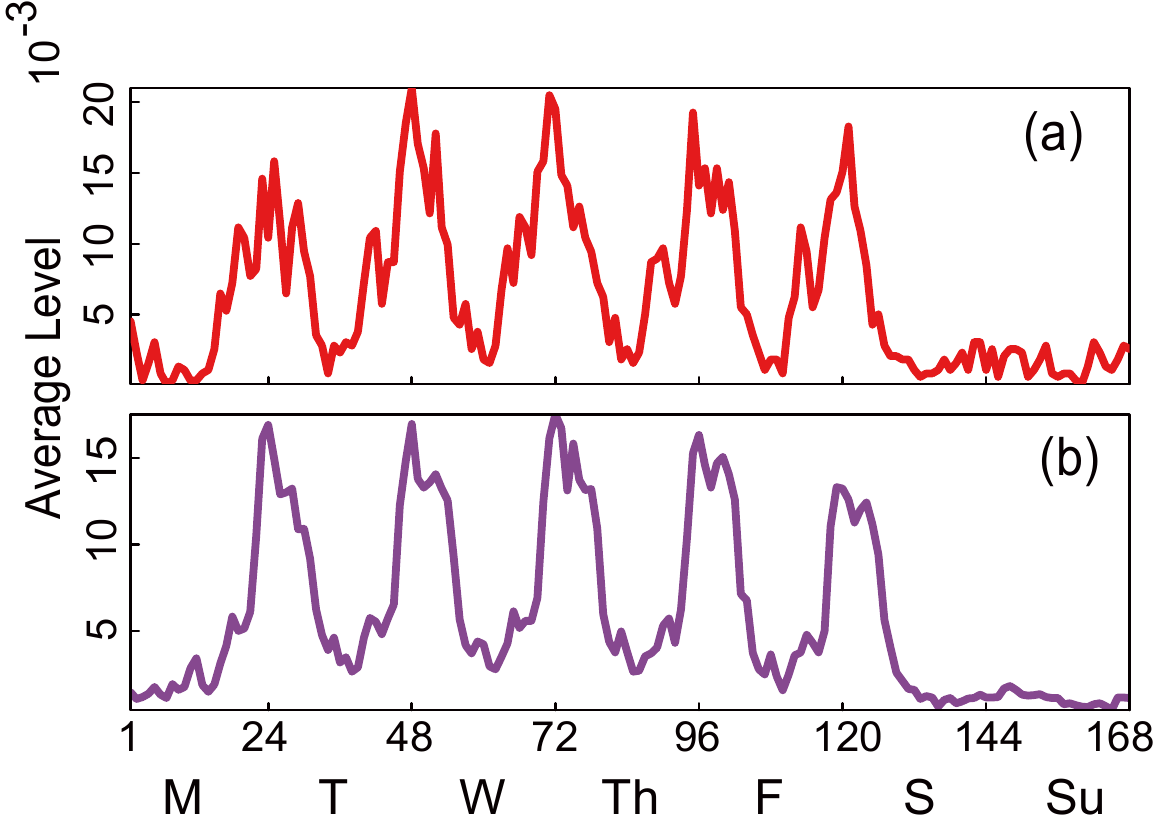}
\caption{Routine activities of developers in the Eclipse project: (a) Average levels of bug-fixing activities within a week and (b) Average levels of bug-tossing activities within a week. The horizontal axis indicates the hours passed since midnight on Monday, from 1 to 168. The vertical axis denotes the average level of bug-fixing (or bug-tossing) activities, which are normalized by the number of fixing (or tossing) events that occur within an hour to the total number of those occurring fixing (or tossing) activities during seven days.}
\label{periodic-activities}
\end{figure}

Although the results of existing approaches are promising, there still exist a few challenges. On the one hand, bug tossing graphs used in previous studies are treated as a class of static graphs, which contain a fixed set of nodes and edges. However, they are dynamic and evolving in the real-world. Moreover, the size of a DCN is growing with an explosion of developer activities in the open-source community. Unfortunately, current methods still have limitations in analyzing large-scale networks efficiently. On the other hand, developer activities are also affected by circadian rhythms in humans \cite{Vitaterna2001Overview}. Fig. \ref{periodic-activities} presents bug-fixing and bug-tossing patterns along the time axis in the Eclipse project, in which developers exhibit distinct behavioral patterns at different hours, workdays, and weekends. The periodic interaction between developers may affect the performance of developer recommendations. Nevertheless, none of the previous studies have taken into account this factor.

Due to the prevalence of large-scale networked systems \cite{Minimizing2016,tang_leveraging_2011,  HuangM19}, graph nodes’ latent representation (or embedding) has been widely investigated. The basic idea of graph representation learning is to learn a low-dimensional vector for each node in a graph. Such a node embedding encodes a graph’s structural information and properties. The low-dimensional vector representation for nodes promotes a variety of network analysis tasks, such as node classification \cite{perozzi_deepwalk_2014,grover_node2vec_2016}, link prediction \cite{ibrahim_link_2015,yu_link_2017}, and recommendation \cite{ying_graph_2018,krishnan_adversarial_2018}. Inspired by the idea of graph representation learning, in this article, we propose a spatial-temporal dynamic graph neural network (ST-DGNN) framework to learn the spatial-temporal features of dynamic DCNs. In addition to a joint random walk (JRWalk) mechanism that samples local topological structure with two types of sampling strategies, this framework contains a graph recurrent convolutional neural network (GRCNN) model with hourly-periodic, daily-periodic, and weekly-periodic components.

To demonstrate the proposed ST-DGNN framework’s effectiveness, we investigate the following three research questions (RQs): 

\begin{itemize}
    \item[(1)]
    Does the JRWalk mechanism capture more diverse and informative information than the state-of-the-art (SOTA) neighborhood sampling methods?
    \item[(2)] 
    Is the GRCNN model more useful for learning spatial-temporal features than the SOTA models in the task of developer attribute prediction (DAP)?
    \item[(3)] 
    Does the proposed framework outperform other baseline approaches in the task of bug fixer prediction (BFP)?
\end{itemize}

In brief, the main contributions of this study are summarized as follows.

\textbf{Joint Random Walks}. We design the JRWalk mechanism, which takes into account both edge importance and degree importance. Many previous random-walk-based studies consider only a single sampling strategy. JRWalk samples the local topological structure of DCNs by leveraging the tossing preference between developers (i.e., edge importance) and developers’ reputation (i.e., degree importance). Such a hybrid strategy can improve traditional random-walk strategies in capturing more diverse and informative information. 

\textbf{Graph Recurrent Convolutional Neural Network}. We construct the GRCNN model with JRWalk to capture a DCN’s spatial-temporal features. For each node sequence obtained by JRWalk, we use a convolutional neural network (CNN) to extract the spatial features. Considering the periodicity of bug-fixing activities, we model the hourly-periodic, daily-periodic, and weekly-periodic dependencies of a DCN by multiple long short-term memory (LSTM) units to learn the temporal features. An attention mechanism is also applied to the LSTM network to capture the dynamic temporal correlations of network data between different time slices.

\textbf{Evaluation}. We compare the ST-DGNN framework with traditional developer recommendation methods and deep-learning-based methods on two real-world, large-scale datasets collected from the Eclipse and Mozilla projects. Experimental results indicate that the proposed framework performs better than all the selected baseline approaches in the BFP task.
         
The rest of this paper is structured as follows. Section \ref{sec:relatedwork} introduces the related work. Section \ref{sec:ProblemDef} formulates the underlying problem investigated in this study. Section \ref{sec:Model} details the ST-DGNN framework and two domain-specific tasks, and Section \ref{sec:Experiments} shows the performance comparison among different approaches on two real-world datasets in the two tasks. Section \ref{sec:Discussion} discusses the GRCNN model variants and the sensitivity analysis on a few critical parameters in our work. Section \ref{sec:Threats} identifies some potential threats to the validity of this study. Finally, Section \ref{sec:Conclusion} concludes this paper.

\section{Related Work} \label{sec:relatedwork}

\subsection{Graph Representation Learning}

It has long been recognized that graphs are a powerful tool to describe natural and human-made systems. With the popularity of graph-structured data streams in real-world applications, many vital tasks related to graphs (e.g., making predictions over nodes and edges) demand effective graph representation learning algorithms to extract meaningful features and patterns. Generally speaking, this study’s related work includes representation learning methods on static graphs and recent deep-learning-based advancements on dynamic graphs.

There are many approaches proposed for graph representation learning. Early studies in this field mainly focused on dimensionality reduction, which decomposes the graph Laplacian or high-order adjacency matrix to produce node representations \cite{belkin_laplacian_2001,cao_grarep_2015,ou_asymmetric_2016}. However, these methods are time-consuming and have low storage efficiency. Inspired by those representation learning models in natural language processing (NLP), DeepWalk \cite{perozzi_deepwalk_2014} generalizes word embeddings from sequences of words to graphs, and leveraged local information obtained from truncated random walks to learn latent representations of nodes in a network. Node2vec \cite{grover_node2vec_2016} extends DeepWalk by adding a combination of the depth-first search and breadth-first search to explore node neighborhoods flexibly and efficiently. LINE \cite{tang2015line} further designs an optimized objective function that preserves the local and global structural information by considering the first-order and second-order proximities to learn network representations. It has been proved to be useful for large-scale graph representations. A few other approaches were also proposed to generate a unified embedding for those attributed networks \cite{chang_heterogeneous_2015,huang_label_2017,huang_graph_2019}.

Representation learning on dynamic graphs has become popular in recent years. Early studies attempted to apply existing typical models for static graphs to evolving networks. For example, Nguyen et al. \cite{nguyen_continuous-time_2018} designed a skip-gram model with a temporal random walk to capture continuous-time dynamic networks’ critical temporal properties. Du et al. \cite{du_dynamic_2018} extended the skip-gram model in a dynamic setting to learn representations of new vertexes and update the original ones. With the great success of deep learning in computer vision and NLP, researchers began to use deep learning techniques to characterize dynamic networks. 

Trivedi et al. \cite{trivedi_know-evolve_2017} proposed a deep evolutionary knowledge network architecture that learns non-linearly evolving entity representations over time. They further designed a two-time scale deep temporal point process model, DyRep \cite{trivedi_dyrep_2019}, to capture the temporal interaction between nodes and the topological evolution of a dynamic graph. Zhou et al. \cite{zhou_dynamic_2018} treated the triadic closure process as a fundamental mechanism in the formation and evolution of dynamic networks. They proposed a model called DynamicTriad to learn representation vectors for each node at different time points. To learn the temporal transition of a network, Goyal et al. \cite{goyal_dyngraph2vec_2020} proposed a method called dyngraph2vec, which extends DyGEM \cite{goyal_dyngem_2018} with a deep architecture composed of dense and recurrent layers. 

Besides, attention mechanisms, which select the information relatively critical to the input information’s current task, have been widely used in various deep learning architectures. For example, Velickovic et al.  \cite{velickovic_graph_2018} designed a graph attention network that specifies different weights to nodes in a neighborhood. Liang et al. \cite{liang_geoman_2018} proposed a multilevel attention network that can adjust the correlations of graphs among multiple time series adaptively. Thekumparampil et al. \cite{thekumparampil_attention-based_2018} presented a novel graph neural network (GNN) in which an attention mechanism allows the network to learn a dynamic and adaptive local summary of the neighborhood. Zhang et al. \cite{zhang_improving_2020} theoretically analyzed the discriminative power of GNNs with attention-based aggregators.

\subsection{Deep-Learning-based Bug Triaging}

Existing literature on bug triaging has proved that automated approaches based on machine learning are useful in recommending appropriate fixers for confirmed bug reports in bug tracking systems. With the rapid development of deep learning, some deep-learning-based methods were recently proposed to improve existing bug triaging approaches based on traditional machine learning techniques. For example, Mani et al. \cite{mani_deeptriage_2019} proposed an attention-based deep bidirectional recurrent neural network (DBRNN-A) model to recommend the most appropriate developers for a given bug report. Lee et al. \cite{LeeHLKJ17} used a CNN that extracts features from bug reports’ textual content to build an automated bug triager. Guo et al. \cite{GuoZYCGLL20} found that developer activities can help distinguish similar bug reports. They presented a CNN-based bug triaging approach that leverages developer activities. Xi et al. \cite{XiYXXL19} proposed an automatic bug triaging approach that adopts a sequence-to-sequence model to jointly learn the features of textual content and the tossing sequence of bug reports. Choquette-Choo et al. \cite{Choquette-ChooS19} exploited developers’ team classes to improve the efficiency of bug assignment. They introduced a dual deep neural network architecture (Dual DNN) for triaging bug reports to both a team and an individual developer. 

Generally speaking, most of the above studies used bug reports’ text features and treated bug triage as a text classification problem. Besides, it is worth noting that a DCN’s topology contains vital information on bug tossing that can improve bug triaging efficiency \cite{jeong_improving_2009}. Recently, a small number of researchers also attempted to leverage hidden features extracted from bug tossing graphs by network representation learning. For example, Huang et al. \cite{HuangM19} considered structural information of a multiplex DCN and then integrated the idea of network embedding to propose an automatic bug triaging approach. Alazzam et al. \cite{AlazzamALK20} proposed a graph-based feature augmentation approach that utilizes graph partitioning based on neighborhood overlap. However, they conducted experiments on static graphs in a single time slice, and features hidden in changing DCNs (i.e., dynamic graphs) were not fully utilized.

In recent years, GNNs have been widely used to extract information about the relationship between nodes in a graph, making many achievements in research areas related to graph analysis. Therefore, we apply the GNN theory to investigate the automatic bug triaging approach in this work to extract high-level features from changing DCNs effectively.

\section{Problem Definition} \label{sec:ProblemDef}

The bug triaging process has two essential tasks: bug fixer prediction and developer attribute prediction. Both of the two tasks need to utilize DCNs’ structure information and their evolution information, which comes down to learning graph representations of changing DCNs. 

In this study, we define a changing DCN as a dynamic graph $G = \{G_1, ..., G_T \}$, i.e., a series of the network’s observed snapshots. Here, $T$ denotes the number of time slices. A DCN at time slice $t$ is then defined as $G_t = (V_t, E_t)$, where $V_t$ and $E_t$ are sets of nodes and links, respectively. For such a network, a node represents a developer, and a link represents the bug tossing relationship. A DCN is, in essence, a directed weighted graph with a weighted adjacency matrix denoted as $\bm{G}_t \in \mathbb{R}^{N \times N}$. For developers $i$ and $j$, if there is no link between them, $w_{ij}$ is zero; otherwise, $w_{ij}$ is larger than zero, and the value indicates the strength of their interacting relationship. 

The goal of dynamic graph representation is to learn a latent representation $\bm{g}_v \in \mathbb{R}^d$ of each node $v \in V_t$ during $T$ time slices. Here, $\bm{g}_v^t$ denotes the latent representation of node $v$ at time slice $t$, and $g_v^t$ represents the local graph structure (i.e., a small sub-graph) centered at $v$ and meanwhile preserve its evolutionary behaviors prior to $t$.

In this article, we use uppercase bold letters (e.g., $\bm{G}$ and $\bm{H}$) to denote matrices, boldface lowercase alphabets (e.g., $\bm{g}$ and $\bm{h}$) to denote vectors, and lowercase alphabets (e.g., $g$ and $h$) to denote scalars. Table \ref{table_notation} presents primary notations and their definitions used in this article.

\begin{table*}[t]
\centering
\caption{Primary notations and definitions used in this article.}
\label{table_notation}
\begin{tabular}{lll}
\hline
Notation & Definition                           &  \\ \hline
$N = |V| \in \mathbb{N}$   & number of nodes in a graph                &  \\
$\bm{G}_t \in \mathbb{R}^{N \times N}$        & weighted adjacency matrix at time slice $t$ &  \\
$\alpha \in \mathbb{R}_{+}$         & probability of determining the next transition strategy for a node &  \\
$l \in \mathbb{N}$         & length of a joint random walk &  \\
$r \in \mathbb{N}$         & number of joint random walks sampled for each node &  \\
$P_{e}(i \rightarrow j)$  & probability of node $i$ jumping to $j$ using the strategy based on edge preference  & \\
$P_{v}(i \rightarrow j)$  & probability of node $i$ jumping to $j$ using the strategy based on node degree preference & \\ 
$\bm{g}_i^t$ & latent representation of node $i$ at time slice $t$  &  \\
$\hat{\bm{y}}$ & prediction result for a given GRCNN component &  \\
$\tilde{\bm{y}}$ & final prediction result for a given task &  \\
\hline
\end{tabular}
\end{table*}

\section{ST-DGNN Framework and Related Tasks} \label{sec:Model}

\begin{figure*}[hbt!]
\centering
\includegraphics[width=4in]{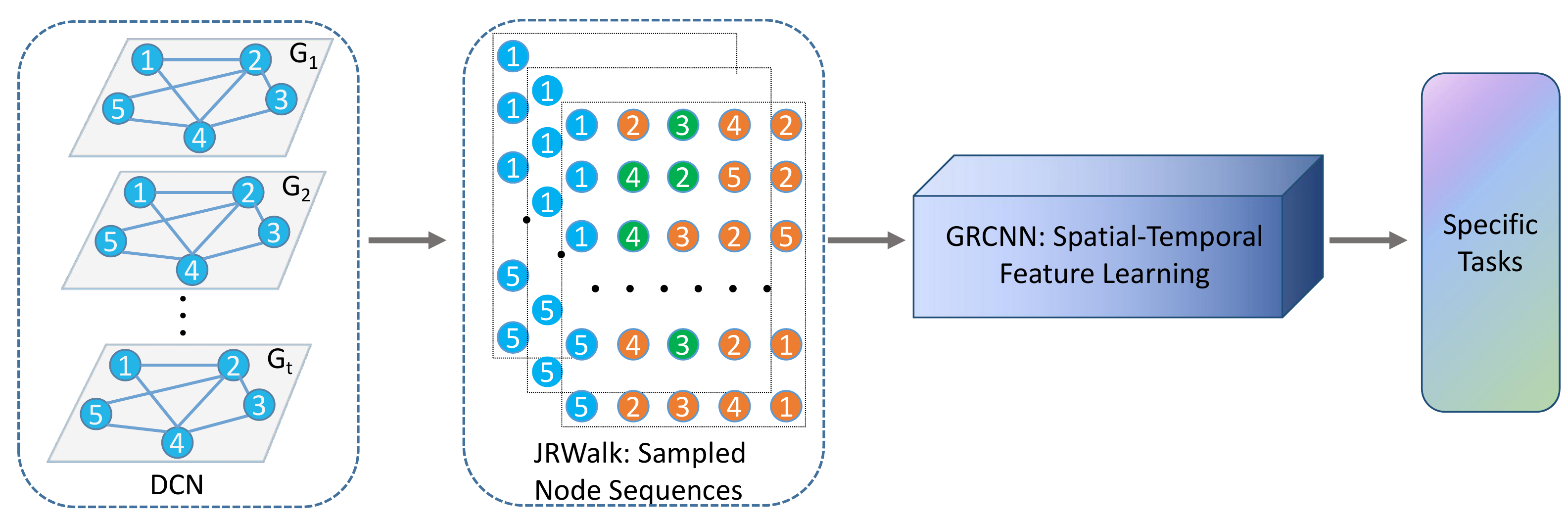}
\caption{The proposed spatial-temporal dynamic graph neural network framework. The input of this framework is a series of DCNs at different time slices. JRWalk processes each DCN and then obtains the network’s node sequences (i.e., walks). GRCNN extracts spatial-temporal features embodied in all the DCNs with walks. Finally, the learned node representations are used for two domain-specific tasks.}
\label{fig_framework}
\end{figure*}

As shown in Fig. \ref{fig_framework}, the ST-DGNN framework has two main modules, i.e., the JRWalk mechanism and the GRCNN model. This framework’s key idea is to model complex interactions between developers by JRWalk and then use GRCNN to learn spatial-temporal features embodied in changing DCNs with walks (i.e., node sequences) obtained by JRWalk. The details of JRwalk and GRCNN are introduced in Subsections \ref{sec-JRW} and \ref{sec-GRCNN}, respectively. 

\subsection{Joint Random Walk} \label{sec-JRW}

Random-walk-based graph representation has been studied for years  \cite{perozzi_deepwalk_2014,grover_node2vec_2016,hamilton_inductive_2017}. Various sampling strategies for nodes in a graph generate different feature representations. In a DCN, the interaction (more specifically, bug tossing) between developers is related closely to developers’ reputation and preference. In this study, we use node degree and edge weight to represent reputation and preference, respectively. Thus, we can bias the random walks in an individual-specific way, which jointly considers both node importance and edge importance to sample neighbors of a given node.

\begin{figure*}[hbt!]
\centering
\includegraphics[width=4in]{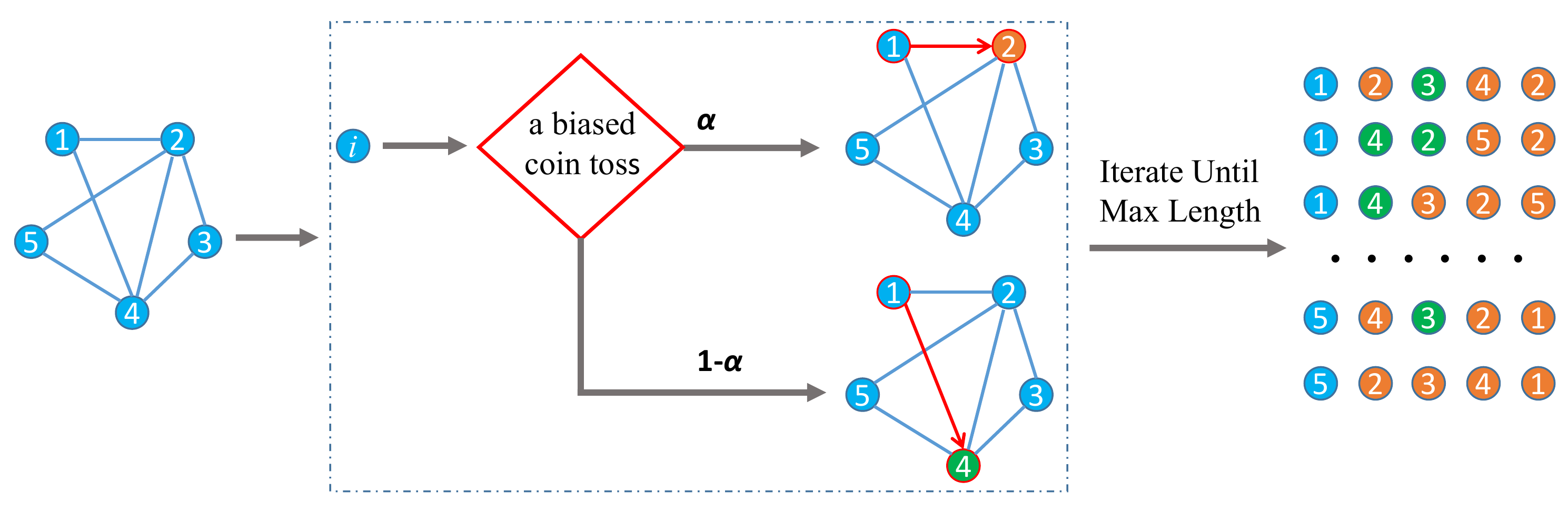}
\caption{An illustration of the joint random walk. A node would jump to its neighboring nodes with probability $\alpha$ by using the edge-preference-based strategy and $1 - \alpha$ by using the degree-based strategy, which increases the diversity and information of the walks. Suppose node $1$ and node $2$ are the current and the next-step node, respectively. Node $1$ walks to node $2$ with probability $P_{e}(1 \rightarrow 2)$ when the coin toss yields head; otherwise, node $1$ jumps to node $4$ with probability $P_{v}(1 \rightarrow 4)$.}
\label{fig_walk}
\end{figure*}

One of our approach’s key issues is how we sample neighbors $N_S(i)$ of node $i$ effectively. To this end, we use JRWalk to make the random walks more informative. As illustrated in Fig. \ref{fig_walk}, we consider a biased coin toss that determines the next transition strategy for a node. The coin comes up head with probability $\alpha$, implying that the hop from one node to the other follows an edge-preference-based strategy.

Suppose $i$ and $j$ denote the current node and the next-step node, respectively. Node $i$ jumps to node $j$ with probability $P_{e}(i \rightarrow j)$. In Eq. \ref{TP}, $w_{ij}$ denotes the weight of an edge between nodes $i$ and $j$.

\begin{equation}
P_{e}(i \rightarrow j) = \frac{w_{ij}}{\sum_{p=1}^{N}w_{ip}}.
\label{TP}
\end{equation}

If the coin turns tail, $i$ walks to $j$ according to a degree-based strategy, with probability $P_{v}(i \rightarrow j)$ defined as

\begin{equation}
P_{v}(i \rightarrow j) = \frac{d_{j}}{\sum_{p=1}^{N}d_{p}},
\label{Deg}
\end{equation}
where $d_p$ denotes the degree of node $p$.

We can transform local topological structures of all nodes in a DCN into informative node sequences by JRWalk. For each node $i \in V_t$, we perform $r$ joint random walks of length $l$, each of which starts from $i$. If the number of nodes in a walk is less than $l$, we conduct the padding operation. Node sequences generated by the joint random walks from start node $i$ is denoted as set $s_i$, and node sequences generated for all nodes in a network are denoted as set $S =\bigcup_{i=1}^{N}s_i$. We illustrate how JRWalk works in Fig.  \ref{fig_walk}. For example, node $1$ walks to node $2$ with probability $P_{e}(1 \rightarrow 2)$ when the coin toss yields head; otherwise, node $1$ jumps to node $4$ with probability $P_{v}(1 \rightarrow 4)$. Orange and green circles denote nodes obtained by the edge-preference-based and the degree-based strategy, respectively. If we set $l = 5$ and $r = 3$ in Fig. \ref{fig_walk}, the set of node sequences generated by JRWalk for node $i$ is $\{\{1, 2, 3, 4, 2\}, \{1, 2, 4, 5, 2\}, \{1, 4, 3, 2, 5\}\}$.

Algorithm \ref{Algorithm 1} presents the generation procedure of the joint random walks from node $i$. Given a network (i.e., a snapshot of the DCN) $G_t = (V_t, E_t)$, we firstly compute the transition probabilities ${P}_{e}$ and ${P}_{v}$ for each node in $G_t$. For each node $i \in V_t$, we sample $r$ node sequences of length $l$, each of which starts from $i$, and then append them to $s_i$, which is the output of Algorithm \ref{Algorithm 1}. In particular, we construct an alias table \cite{Devroye86} to sample nodes from the transition matrices $\bm{P}_{e} \in \mathbb{R}^{N \times N}$ and $\bm{P}_{v} \in \mathbb{R}^{N \times N}$, making sure that the jumping between two nodes can be done efficiently in $O(1)$. This process can reduce the computation complexity when simulating the joint random walks in a large-scale network.

\begin{algorithm}[ht]
\caption{JRWalk($G_t$,$i$,$r$,$l$,$\alpha$)}
\label{Algorithm 1}
\begin{algorithmic}[1]
\renewcommand{\algorithmicrequire}{\textbf{Input:}}
\renewcommand{\algorithmicensure}{\textbf{Output:}}
\REQUIRE graph $G_t$, start node $i$, walks per node $r$, walk length $l$, and probability $\alpha$
\ENSURE  set of joint random walks $s_i$ for node $i$
\STATE Get direct neighbors $N_S(i)$ of node $i$; 
\FOR {$j = 1$ to $r$}
\STATE Initialize $i$ to $s_{ij}$; 
\FOR {$k = 2$ to $l$}
\IF{$\alpha >$ rand()}
\STATE Sample node $n_k$ from $N_S(i)$ with probability $P_{e}$ using an alias table; 
\ELSE
\STATE Sample node $n_k$ from $N_S(i)$ with probability $P_{v}$ using an alias table; 
\ENDIF
\STATE Append $n_k$ to $s_{ij}$;
\ENDFOR
\STATE Append $s_{ij}$ to $s_i$;
\ENDFOR
\RETURN $s_i$;
\end{algorithmic}
\end{algorithm}

\subsection{Graph Recurrent Convolutional Neural Network} \label{sec-GRCNN}

\begin{figure*}[hbt!]
\centering
\includegraphics[width=6in]{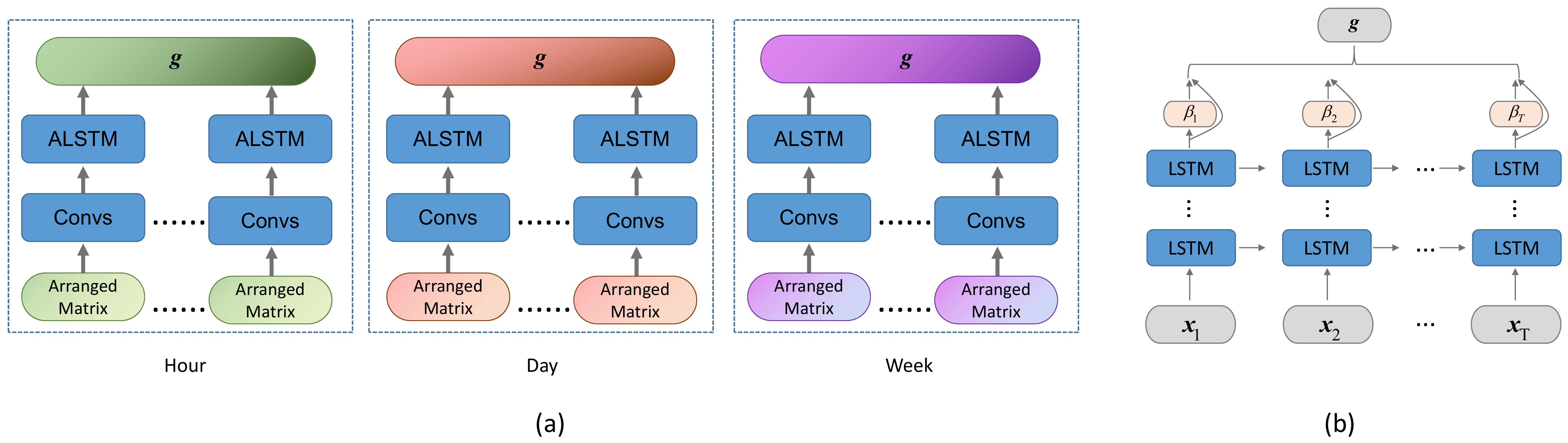}
\caption{The GRCNN model architecture: (a) Overview of the GRCNN model and (b) Details of the ALSTM module. Convs: Convolution operation in a CNN module; ALSTM: Attention-based Long Short-term Memory Module.}
\label{fig_model}
\end{figure*}

After obtaining the set of node sequences for each snapshot of a DCN, we utilize an embedding method to learn a latent representation of each node. In this study, we use GRCNN to perform graph embedding on the learned node sequences. As shown in Fig. \ref{fig_model}(a), GRCNN has three components with the same structure: hourly-periodic, daily-periodic, and weekly-periodic components. Each component’s goal is to capture the spatial-temporal features of nodes in a snapshot of the DCN on the corresponding time scale. A GRCNN component consists of two necessary parts: a CNN module and an attention-based LSTM (ALSTM) module. The CNN module is used to exploit the high-level spatial features of nodes in each snapshot. The ALSTM module is employed to adaptively capture critical temporal dependencies among snapshots at different time slices. The details of a GRCNN component’s two parts are introduced as follows.

\subsubsection{Convolution in Spatial Dimension}

For an input graph $G_t$ at time slice $t$, JRWalk generates unified walking sequences of length $l$ for each node in the graph. We then map each of $l$ nodes in a walking sequence into a $a_v$-dimensional vector. Therefore, we can obtain one tensor ($N$, $l$, $r$, $a_v$) after performing the joint random walk for a graph. This tensor can be reshaped to a three-dimensional tensor ($N$, $lr$, $a_v$). Here, $\bm{a}$ denotes a one-hot encoded vector for each node (or node attributes if available), and $a_v$ represents the number of node attributes. With this approach, the convolution operations on arranged matrices in each CNN module are conducted in the same way that CNNs are applied to two-dimensional lattices such as images. The convolution operator is defined as

\begin{equation}
\bm{x}_{j}^{l} = relu(\sum_{i \in M_{j}} \bm{x}_{i}^{l-1} \cdot \bm{W}_{ij}^{l} + \bm{b}_{j}^{l}),
\label{convs}
\end{equation}
where $\bm{x}_{j}^{l}$ is the $j$-th feature map of the $l$-th layer in the CNN module, $\bm{W}_{ij}^{l}$ is the weight matrix of the $j$-th convolutional filter from $\bm{x}_{i}^{l-1}$ to $\bm{x}_j^{l}$, $M_{j}$ is the number of feature maps in layer $l-1$, and $\bm{b}_{j}^{l}$ is the bias term for the output $\bm{x}_{j}^{l}$. Note that $relu$ is the rectified linear unit (ReLU) function that ranges from 0 to 1. Here, we use it to avoid the over fitting problem.

\subsubsection{Attention-based LSTM Module}

There exist correlations between nodes within a walking sequence in the time dimension. Moreover, such correlations may vary from time to time because the DCN is changing. Because LSTM units can capture long-range dependencies in a sequential pattern, we employ the ALSTM modules (see Fig. \ref{fig_model}(b)) to capture these nodes’ high-level temporal features at different time slices. In particular, we utilize an attention mechanism to weigh hidden states generated by LSTM units adaptively. Formulas for LSTM units in each ALSTM module are described as follows:

\begin{equation}
\bm{f}_{t} = sigmoid(\bm{W}_{f} \bm{x}_t + \bm{U}_{f} \bm{h}_{t-1} + \bm{b}_f),
\label{LSTM-f}
\end{equation}

\begin{equation}
\bm{i}_{t} = sigmoid(\bm{W}_{i} \bm{x}_t + \bm{U}_{i} \bm{c}_{t-1} + \bm{b}_i),
\label{LSTM-i}
\end{equation}

\begin{equation}
\bm{o}_{t} = sigmoid(\bm{W}_{o} \bm{x}_t + \bm{U}_{o} \bm{c}_{t-1} + \bm{b}_o),
\label{LSTM-o}
\end{equation}
where $\bm{x}_t$ is the current feature map generated by the CNN module at time slice $t$, $\bm{f}_t$, $\bm{i}_t$, and $\bm{o}_t$ are the forgotten gate, input gate, and output gate, respectively, $\bm{W}_*$ and $\bm{U}_*$ are the input weight matrix and the recurrent weight matrix, respectively, and $\bm{b}_*$ is the bias term.

\begin{equation}
\tilde{\bm{c}}_{t} = tanh(\bm{W}_{c} \bm{x}_t + \bm{U}_{c} \bm{h}_{t-1} + \bm{b}_c),
\label{LSTM-c}
\end{equation}

\begin{equation}
\bm{c}_{t} = \bm{f}_t \odot \bm{c}_{t-1} + \bm{i}_t \odot \tilde{\bm{c}}_{t},
\label{LSTM-c2}
\end{equation}

\begin{equation}
\bm{h}_{t} = \bm{o}_t \odot tanh(\bm{c}_t),
\label{LSTM-h}
\end{equation}
where $\bm{h}_t$ is the output hidden state, the operator $\odot$ denotes component-wise multiplication, and $\tilde{\bm{c}}_t$ is a candidate hidden state calculated based on the current input and the previous hidden state. Here, $\tilde{\bm{c}}_t$ is updated to cell state $\bm{c}_{t}$ by combining previous memory cells.

When training these LSTM units, we dynamically assign different weights to the output hidden states at different time slices by leveraging an attention mechanism (more specifically, an additive attention \cite{BahdanauCB14}). In this study, we empirically use a softmax function to calculate the importance value $\beta_t$ for $\bm{h}_t$, defined as follows:

\begin{equation} \label{attention-beta}
\beta_t = \frac{exp(e_t)}{\sum_{t} exp(e_t)},
\end{equation}
where $e_t$ is the non-normalized scalar score obtained by a non-linear transformation of $\bm{h}_t$, formally defined as

\begin{equation} \label{attention-e}
e_t = \bm{v}_e^T tanh(\bm{W}_e \bm{h}_t + \bm{b}_e),
\end{equation}
where $\bm{v}_e$ denotes the vector of a learnable parameter.

The latent representation (or high-level spatial-temporal features) of each node $i$ in a changing DCN is then calculated as a weighted sum of $\bm{h}_t$, defined as below:

\begin{equation} \label{attention-sum}
\bm{g}_i = \sum_{t} \bm{g}_i^t = \sum_{t} \beta_t \bm{h}_t.
\end{equation}

\subsection{Domain-specific Tasks} \label{sec:Task}

\begin{figure}[hbt!]
\centering
\includegraphics[width=3in]{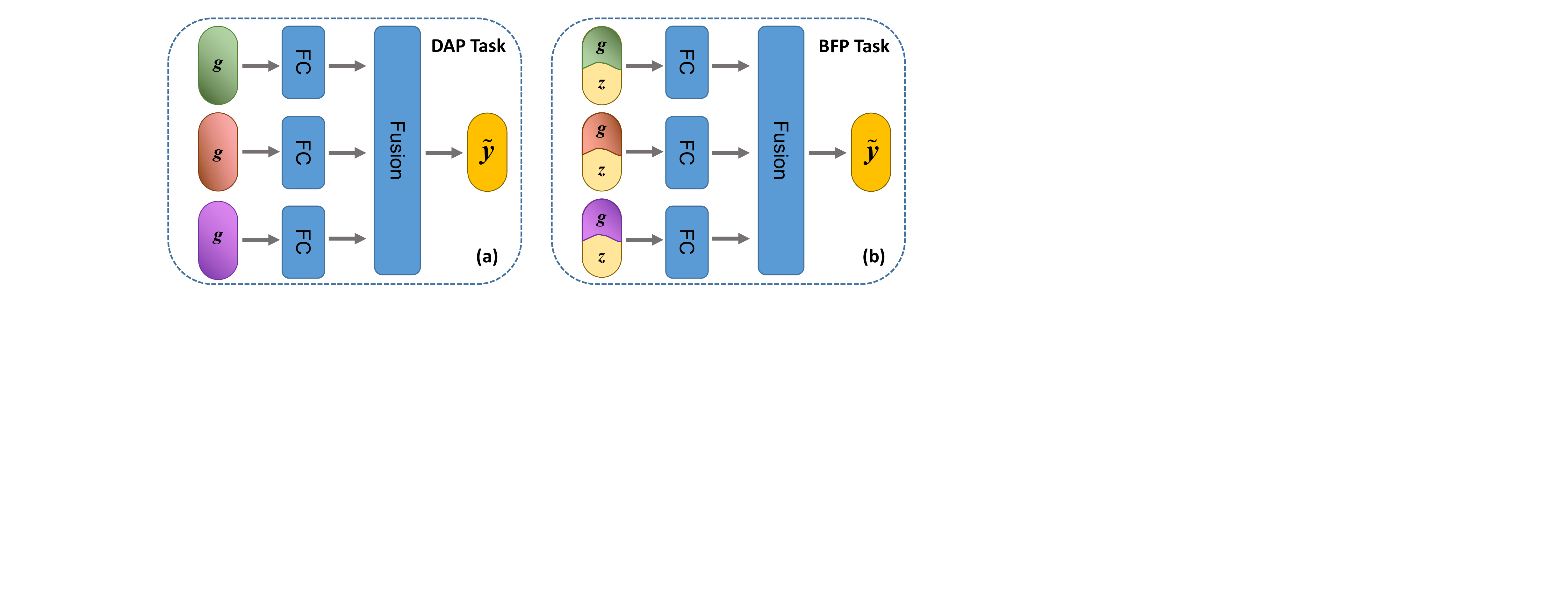}
\caption{Two domain-specific tasks of node classification: (a) the DAP task and (b) the BFP task. FC and Fusion denote the full-connected layer and the fusion operation, respectively.}
\label{fig_task}
\end{figure}

After learning a dynamic graph representation of a changing DCN, we need to perform specific tasks to evaluate the performance of our approach. As shown in Fig. \ref{fig_task}, there are two domain-specific tasks: developer attribute prediction (DAP) and bug fixer prediction (BFP). The DAP task’s goal is to predict a developer’s specific attributes (e.g., expertise and workspace), which plays an important role in improving the accuracy of bug triaging. We perform this task to answer the second research question. The BFP task’s goal is to predict a developer who can fix the target bug. We perform this task to answer the third research question.

Although the two tasks require different input vectors, they are, in essence, a node classification task. Therefore, we can handle the two tasks in a unified way. More specifically, the BFP task has the same operations as the DAP task, including a full-connected (FC) layer, the fusion operation, and a softmax function. Here, we define a generic input $\bm{\mu}$ for the two tasks to simplify the mathematical formulas for the operations in Fig. \ref{fig_task}.

\begin{equation} \label{mu}
\bm{\mu}= \left\{\begin{matrix}
 \bm{g}_i& for~the~DAP~task& \\ 
 concat(\bm{g}_i,\bm{z}_r)& for~the~BFP~task& 
\end{matrix}\right.
\end{equation}

If we perform the DAP task (see Fig. \ref{fig_task}(a)), in Eq. \ref{mu}, $\bm{\mu}$ is equal to  $\bm{g}_i$, which is the spatial-temporal representation of  developer $i$ obtained by GRCNN. If we perform the BFP task (see Fig. \ref{fig_task}(b)), $\bm{\mu}$ is equal to the concatenation of $\bm{g}_i$ and $\bm{z}_r$. Here, $\bm{z}_r$ denotes the text features of each bug report $r$ obtained by latent Dirichlet allocation (LDA).

\subsubsection{Task Execution}

Suppose $\hat{\bm{y}}$ denotes the prediction result of each time-scale component. It is defined as a function of $\bm{\mu}$ in this study. More specifically, we use a FC layer with a softmax function to output $\hat{\bm{y}}$ (see Eq. \ref{attention-fc}).

\begin{equation} \label{attention-fc}
\hat{\bm{y}} = softmax(\bm{W}_{y}^T \bm{\mu} + \bm{b}_y),
\end{equation}
where $\bm{W}_y$ and $\bm{b}_y$ are the FC layer’s parameters.

Because DCNs are changing over time, time scales such as day and week may affect the two tasks’ results. In the ST-DGNN framework, the outputs of GRCNN’s three components (i.e., $\hat{\bm{y}}_h$, $\hat{\bm{y}}_d$, and $\hat{\bm{y}}_w$) are then fused to generate the final prediction result, defined as

\begin{equation} \label{fusion}
\tilde{\bm{y}} = sigmoid(\bm{w}_h  \odot \hat{\bm{y}}_{h} + \bm{w}_d  \odot \hat{\bm{y}}_{d} + \bm{w}_w \odot \hat{\bm{y}}_{w}),
\end{equation}
where $\bm{w}_h$, $\bm{w}_d$, and $\bm{w}_w$ are weight vectors that reflect the three components’ impacts on the prediction result.

\subsubsection{Model Training}

For the DAP task, we define a training instance as $I_i^{t} = <s_i^{t}, \bm{y}_i> $ for node $i$ at time slice $t$. Here, $s_i^{t}$ is a node sequence set generated by JRWalk at $t$, and $\bm{y}_i$ is this node’s actual label denoted by a one-hot vector. We construct the training set $D^a$ (for a dynamic graph $G$) composed of training samples $D_i^a = \{I_i^{1}, I_i^{2}, ......, I_i^{T}$\} to train a GRCNN model with a supervised setting.

For the BFP task, a training instance $\ddot{I}_r^{t} = <s_{h(r)}^{t}, \bm{z}_r, \bm{y}_r>$ at time slice $t$ contains both network structure information and textual content of a bug report. Here, $s_{h(r)}^{t}$ is the network representation of the corresponding developer who is currently dealing with $r$, $h(.)$ is a function that maps $r$ to its current holder, and $\bm{y}_r$ is the actual fixer for bug report $r$, deonted by a one-hot vector. The training set $D^f = \{ D^f_1, D^f_2,..., D^f_R\}$ consists of $R$ training samples, and each training sample $D_r^f = \{\ddot{I}_r^{1}, \ddot{I}_r^{2}, ......, \ddot{I}_r^{T}$\} includes $T$ training instances.

For simplicity, we present a unified objective function (or loss function) form for the two tasks, which is defined based on the cross-entropy loss as

\begin{equation} \label{loss}
J = - \sum_{j \in D} \bm{y}_j log\tilde{\bm{y}_j}.
\end{equation}

Algorithm \ref{Algorithm 2} presents GRCNN’s general training process for the two tasks. Given the training set $D$ ($D^a$ or $D^f$), we train GRCNN based on the above loss function (see Eq. \ref{loss}). The parameter set $\Theta$ (i.e., the output of Algorithm \ref{Algorithm 2}) embraces the weight matrices, bias terms, and other parameters defined in Eq. \ref{convs} $\sim$ Eq. \ref{fusion}, and it is updated during the whole training process. After training GRCNN over $maxepoch$ epochs, we obtain a trained GRCNN model and use it for the corresponding prediction task.

Note that GRCNN can be also trained with an unsupervised or semi-supervised setting because of the property inherited from graph convolutional networks (GCNs) \cite{hamilton_inductive_2017,KipfW17}. For example, we can replace the loss function $J$ with a negative-sampling-based unsupervised objective function to solve an unsupervised learning task \cite{hamilton_inductive_2017,tang2015line}.

\begin{algorithm}[ht]
\caption{Training Algorithm of GRCNN}
\label{Algorithm 2}
\begin{algorithmic}[1]
\renewcommand{\algorithmicrequire}{\textbf{Input:}}
\renewcommand{\algorithmicensure}{\textbf{Output:}}
\REQUIRE training set $D$ ($D^a$ or $D^f$) and number of epochs $maxepoch$
\ENSURE  parameter set $\Theta$ 
\FOR {$epoch$ = 1 to $maxepoch$}
\STATE Map each node in $D$ into a $a_v$-dimensional vector;
\STATE Perform the convolution operation using Eq. \ref{convs};
\STATE Feed the extracted feature maps to the ALSTM modules and use Eq. \ref{LSTM-f} $\sim$ Eq. \ref{LSTM-h} for forward propagation;
\STATE Compute $\hat{\bm{y}}_h$, $\hat{\bm{y}}_d$, and $\hat{\bm{y}}_w$ using Eq. \ref{attention-fc};
\STATE Compute the prediction result $\tilde{\bm{y}}$ using Eq. \ref{fusion};
\STATE Update $\Theta$ to minimize the loss function in Eq. \ref{loss};
\ENDFOR
\RETURN $\Theta$;
\end{algorithmic}
\end{algorithm}

\section{Experiments and Result Analysis} \label{sec:Experiments}

In this section, we design several comparative experiments to answer the three research questions introduced in Section \ref{sec:introduction} and evaluate our approach’s performance on two real-world, large-scale DCNs.

\subsection{Datasets}

The Eclipse and Mozilla datasets were collected from the Bugzilla platform, a popular bug tracking system for open-source software projects. Each software bug has a structured bug report in this platform. A bug report’s status changes in its whole life cycle. Any bug report, which is not closed and open for fixing, is not suitable to be a training sample. Therefore, we selected only those bug reports whose “Status” and “Resolution” were labeled with “Verified” and “Fixed,” respectively. Finally, we collected 200,000 bug reports (record no. from 1 to 357553) and 220,000 bug reports (record no. from 1 to 660036) from the Eclipse project and the Mozilla project, respectively. A summary of the statistics for the two datasets that are available on GitHub\footnote{https://github.com/ssea-lab/BugTriage/tree/master/raw data} is described as follows.

\textbf{Eclipse:} The Eclipse dataset consists of 200,000 bug reports that have been resolved from Oct. 2001 to Nov. 2011. This dataset contains 3,893 developers who participated in 81 components of the Eclipse project. Each developer’s attribute label is defined as a set of components (i.e., workspace) in which the developer has participated. The DCN derived from this dataset is directed and weighted.

\textbf{Mozilla:} The Mozilla dataset consists of 220,000 bug reports that have also been resolved from Oct. 2001 to Nov. 2011. This dataset contains 7,936 developers involved in 47 components of the Mozilla project. Like the Eclipse dataset, we also constructed a weighted directed DCN for this dataset.

\subsection{Preprocessing}

We removed some inactive developers, who fixed no more than five bugs each year, to alleviate the data sparsity problem. According to the “History” records in the original dataset, tossing interactions between developers were extracted to construct DCNs. Note that we deleted any bug report that has no definite fixer in the “History” attribute. There exists a directed edge from node $i$ to $j$ if the corresponding developer $i$ tosses a bug to $j$. The edge weight indicates the frequency of bug tossed from developer $i$ to $j$. Besides, we took the “Summary,” “Description,” and “Comment” attributes of a bug report together as the bug’s free-form textual content. Then, we used our previous method for textual content processing \cite{wu_empirical_2018} to obtain each developer’s features, such as expertise and interest.

\subsection{Baseline Methods}

Three categories of baseline approaches were involved in our experiments to answer the three research questions. Table \ref{table:baselines} summarizes correlations of these baseline methods to the research questions.

\begin{table*}[]
\centering
\caption{Baseline approaches used in the experiments.}
\label{table:baselines}
\begin{tabular}{lllc}
\hline
Research Question & Task & Approach (publication year) & Dynamic Network           \\ \hline
RQ1 for JRWalk    & DAP  & DeepWalk (2014), LINE (2015), Node2Vec (2016), and TNE (2017)  & $\times$                  \\
RQ2 for GRCNN     & DAP  & \tabincell{l} {DynamicTriad (2018), STGCN (2018), ASTGNN (2019), GCLSTM (2018), \\ AGCLSTM (2019), DCRNN (2018), and DGCNN (2019)} & \checkmark \\
RQ3 for ST-DGNN   & BFP  & BOW+SVM (2006), TGN (2009), DF+SVM (2018), DeepTriage (2019), and ITriage (2019) & \checkmark \\ \hline
\end{tabular}
\end{table*}

\subsubsection{Baselines for RQ1}

We selected four commonly-used node embedding approaches, i.e., DeepWalk, LINE (short for large-scale information network embedding), Node2Vec, and TNE (short for temporal network embedding), to verify the effectiveness of JRWalk.

\textbf{DeepWalk}\cite{perozzi_deepwalk_2014}: It is a skip-gram-based graph embedding method that employs truncated random walks on a network and obtains a node’s corresponding embedding with the NLP modeling technique word2vec.

\textbf{LINE}\cite{tang2015line}: It is an improved method based on DeepWalk, which preserves both first and second-order proximities of nodes. Besides, LINE is one of the SOTA embedding methods for large-scale networks and has been widely applied to different types of networks.

\textbf{Node2Vec}\cite{grover_node2vec_2016}: Following the intuition that random walks through a network can be regarded as sentences in a corpus, Grover et al. believed that the flexibility in exploring neighborhoods is the key to learning richer representations of nodes in a graph. Therefore, Node2Vec combines the depth-first search and breadth-first search to embed a network by mapping nodes from a high-dimensional space into a lower-dimensional vector space.

\textbf{TNE}\cite{zhu_scalable_2017}: It is an embedding approach for dynamic networks, which can model network evolution as a Markov process and learn the embedding vectors of nodes based on matrix factorization. Note that we report the best result of this approach with $\lambda = 0.01$.

\subsubsection{Baselines for RQ2}

We compared GRCNN with seven SOTA models for dynamic graphs, including DynamicTriad, STGCN (short for spatial-temporal GCN), ASTGNN (short for attention-based STGCN), GCLSTM (short for graph convolutional LSTM), AGCLSTM (short for attention enhanced GCLSTM), DCRNN (short for diffusion convolutional recurrent neural network), and DGCNN (short for dynamic spatial-temporal graph CNN), to test whether it can outperform the baseline models.

\textbf{DynamicTriad}\cite{zhou_dynamic_2018-1}: It preserves structural information and evolution patterns of a network by imposing triads, which are basic network blocks composed of three vertices. This triadic closure process is a fundamental mechanism in the formation and evolution of a network, thereby making DynamicTriad able to capture the network dynamics and learn representation vectors for each node at different time slices. Note that we report the best result of this approach with $\beta_0 = 0.1$ and $\beta_1 = 10$.

\textbf{STGCN}\cite{yan_spatial_2018}: It designs multi-scale convolutional filters, each of which is composed of receptive field computation and signal mapping, to encode dynamic graphs. Then, it recursively performs on structured graph data of the temporal and spatial dimensions.

\textbf{ASTGCN}\cite{guo_attention_2019}: It is an extension of STGCN. In particular, attention mechanisms are utilized to model the complex spatial correlations between different locations and capture the temporal correlations between different time slices.

\textbf{GCLSTM}\cite{Zhiyong_highorder}: It is a hybrid network architecture that combines GCNs and recurrent neural networks (RNNs). In this model, the graph convolution operation is introduced to LSTM units at each time step for the input-to-state and state-to-state transitions. Therefore, it can model the action sequence by learning spatiotemporal features automatically.

\textbf{AGCLSTM}\cite{ChenyangSiC0WT19}: It was proposed for human action recognition. As an attention enhanced GCLSTM model, AGCLSTM can capture discriminative features in spatial configuration and temporal dynamics. Besides, it can mine the co-occurrence relationship between the spatial and temporal domains. More specifically, an attention mechanism is employed to enhance key nodes’ features, which improves their spatiotemporal expressions.

\textbf{DCRNN}\cite{LiYS018}: By incorporating spatial and temporal dependency in the traffic flow, DCRNN attempts to address the challenges of complex spatial dependency, non-linear temporal dynamics, and the inherent difficulty of long-term forecasting in traffic flow forecasting. More specifically, this model captures spatial dependency by using bi-directional random walks on a graph and temporal dependency by using the encoder-decoder architecture with scheduled sampling.

\textbf{DGCNN}\cite{diao_dynamic_2019}: It models the spatial dependencies between nodes in a graph with a Laplacian matrix based on the distance between nodes. Since the spatial dependencies change over time in many application scenarios, the core of DGCNN is to find the change of the Laplacian matrix with a dynamic Laplacian matrix estimator.

\subsubsection{Baselines for RQ3}

We applied the proposed framework to bug fixer recommendation and compared it with five SOTA or typical approaches in the field of automated bug triaging, including BOW+SVM, TGN (short for tossing graph model), DF+SVM, DeepTriage, and ITriage.

\textbf{BOW+SVM} \cite{Anvik2006}: It uses the term-frequency-based bag-of-words (BOW) model to represent the text features of bug reports and trains a support vector machine (SVM) classifier to learn the types of bugs that each developer has resolved. The trained SVM classifier then recommends a few possible developers for a new bug report.

\textbf{TGN} \cite{jeong_improving_2009}: To the best of our knowledge, TGN is the first approach that leverages DCNs’ structure information in bug triaging. It is a two-stage approach. In the first stage, TGN predicts possible developers for a given bug report by using machine learning algorithms; in the second stage, it introduces a tossing graph model based on Markov chains to improve the prediction result. Due to its desirable quality, this approach has been used as a baseline by many follow-up studies.

\textbf{DF+SVM} \cite{wu_empirical_2018}: It has been recognized that developer features can affect bug triaging accuracy. Previous studies recommend appropriate developers for a given bug report by feeding hand-crafted developer features to a machine learning algorithm like decision trees and Naïve Bayes. On the one hand, our previous work empirically identified a few critical factors (i.e., theme, product, component) among frequently-used developer features from the “History” records of fixed bug reports \cite{wu_empirical_2018}. On the other hand, many empirical studies indicate that SVM models perform better than other classification algorithms in the developer recommendation task. Therefore, DF+SVM denotes a machine-learning-based method that uses an SVM classifier to predict appropriate developers according to developers’ three critical factors mentioned above.

\textbf{DeepTriage} \cite{mani_deeptriage_2019}: It is a deep-learning-based approach that uses an attention-based deep bi-directional RNN model to learn syntactic and semantic features from long word sequences. An attention mechanism enables this approach to learn better the context representation of a bug from the corresponding long-text bug report.

\textbf{ITriage} \cite{XiYXXL19}: It is also a deep-learning-based approach that consists of two modules: the feature learning model and the fixer recommendation model. ITriage adopts a sequence-to-sequence model to jointly learn the text features from a bug report’s content and the routing features from tossing sequences. Then, a classification model is used to integrate all the features from the textual content, metadata, and tossing sequences to predict bug fixers. 

\subsection{Parameter Settings}

All of the experiments in this study were conducted on a workstation with a 10-core Intel Xeon CPU (E5-2660 v3, 2.60 GHz), 512GB RAM, and two NVIDIA Tesla V100 GPUs. All the baselines’ parameters were set to the configurations in their original papers unless otherwise mentioned. We implemented the proposed framework based on the Pytorch framework\footnote{https://pytorch.org/}, and the operating system and programming language were CentOS 7.0 and Python 3.5, respectively. A few necessary parameters of ST-DGNN are introduced as follows.

For the JRWalk mechanism, the walk length $l$ and the number of walks $r$ were set to 90 and 35, respectively. In Section \ref{sec:Discussion}, we further analyzed the two parameters’ sensitivity.

For the convolution operation in GRCNN, we employed a two-layer CNN to learn spatial features of the input graph at each time slice. In the first convolutional layer, we applied $K_1$ convolution kernels with size $1 \times 3 \times a_v$ to the input tensor ($n$, $lr$, $a_v$) and obtained a $n \times (lr-2) \times K_1$ feature map. The feature map was reshaped and then fed to the second convolutional layer. We used $K_2$ convolution kernels with size $1 \times 3 \times K_1$ to the input feature map in the second convolutional layer. Finally, we got a new feature map ($n \times (lr-4) \times K_2$) from the second convolutional layer. In our experiments, $K_1$ and $K_2$ were set to 64.

For the ALSTM modules in GRCNN, the number of hidden units in each LSTM unit was set to 256. Therefore, $\bm{h}_t$ belongs to $\mathbb{R}^{256}$. In the training process, the dropout rate was set to 0.5 to avoid the over-fitting problem, and we used a fixed learning rate of 0.01. Moreover, we trained GRCNN with the Adam optimizer, and the maximum epoch number was set to 100. The training process of each approach ended when it achieved the best performance on the validation set. Finally, we obtained the latent representation of node $i$, $\bm{g}_i \in \mathbb{R}^{256}$.

We used a two-layer FC with a softmax function as the task classifier for each GRCNN component, and the number of neurons in the two layers was set to 2,048. In the BFP task, we used LDA to extract text features of bug reports. The number of topics was set to 200, and the prior of document topic distribution and the prior of topic word distribution were set to 1/200. 

For more details of the implementation code, please refer to https://github.com/ssea-lab/BugTriage/tree/master/GRCNN.

\subsection{Evaluation}

Because it is hard to evaluate representation learning results obtained by different approaches directly, we perform the DAP and BFP tasks on dynamic DCNs to assess them in this study. To keep the experiments as fair as possible, the Monte Carlo Cross-Validation approach, preserving the percentage of samples for each class, is employed to select the training and test sets. It randomly selects (without replacement) some fraction of a dataset to construct a training set and leaves the rest of the samples as the test set. In our experiments, the original training set was further split into the training and validation sets to tune model hyper-parameters across all approaches. In each experimental dataset, we set aside 10\% of the data for validation, and the rest was split for training and testing, including three data partition schemes. More specifically, the percentages of training data (or test data) were set to 30\% (60\%), 50\% (40\%), and 70\% (20\%), respectively. After repeating the Monte Carlo Cross-Validation approach ten times, we calculated the average performance regarding accuracy (also known as hit rate) to obtain robust estimations for all the methods under discussion. Besides, we also used early-stop to avoid the over-fitting problem in the experiments.

\subsection{Comparison and Result Analysis}

\subsubsection{Results of Different Random Walk Mechanisms} \label{sub_s}

To answer the first research question, we compared JRWalk with four mainstream node embedding approaches (i.e., DeepWalk, Node2Vec, LINE, and TNE) on the Eclipse and Mozilla datasets. We used different amounts of labeled data for training to test the five approaches’ performance in the DAP task. Table \ref{table:node_embedding} presents their results (mean $\pm$ standard deviation) in terms of accuracy. Note that the best result is highlighted in bold.

As shown in Table \ref{table:node_embedding}, LINE performs the best among the four baseline approaches on the two datasets. Although JRWalk beat LINE by a narrow margin on the Eclipse dataset, its accuracy was about 10\% higher than that of LINE on the Mozilla dataset when selecting less training data. These observations indicate that JRWalk can capture nodes’ more useful information by considering both node importance and edge importance.

\begin{table*}[]
\centering
\caption{Results of Different Random Walk Mechanisms (mean $\pm$ s. d.).}
\label{table:node_embedding}
\begin{tabular}{cclllclll}
\hline
\multirow{2}{*}{Approach} & \multicolumn{3}{c}{Eclipse}                                              & \multicolumn{1}{c}{}  & \multicolumn{3}{c}{Mozilla}                                              & \multicolumn{1}{c}{}  \\ \cline{2-9}
                          &                      & \multicolumn{1}{c}{40\%} & \multicolumn{1}{c}{60\%} & \multicolumn{1}{c}{80\%} &                      & \multicolumn{1}{c}{40\%} & \multicolumn{1}{c}{60\%} & \multicolumn{1}{c}{80\%} \\ \cline{1-9}
DeepWalk                  &                      & 0.409$\pm$0.021 & 0.475$\pm$0.014 & 0.520$\pm$0.009 &                      & 0.321$\pm$0.112 & 0.363$\pm$0.028 & 0.443$\pm$0.026 \\
Node2Vec                  &                      & 0.409$\pm$0.017 & 0.477$\pm$0.013 & 0.521$\pm$0.016 &                      & 0.322$\pm$0.024 & 0.364$\pm$0.018 & 0.466$\pm$0.015 \\
LINE                      &                      & 0.413$\pm$0.006 & 0.491$\pm$0.005 & 0.532$\pm$0.002 &                      & 0.323$\pm$0.014 & 0.370$\pm$0.009 & 0.481$\pm$0.012 \\
TNE                       & \multicolumn{1}{l}{} & 0.306$\pm$0.016 & 0.359$\pm$0.011 & 0.384$\pm$0.014 & \multicolumn{1}{l}{} & 0.286$\pm$0.032 & 0.301$\pm$0.013 & 0.352$\pm$0.014 \\
JRWalk                    &                      & \textbf{0.415$\pm$0.011} & \textbf{0.497$\pm$0.008} & \textbf{0.536$\pm$0.006} &                      & \textbf{0.357$\pm$0.016} & \textbf{0.408$\pm$0.007} & \textbf{0.491$\pm$0.008} \\ \hline
\end{tabular}
\end{table*}

\subsubsection{Results of Different GNN Models} \label{sub_gnn}

To explore the second research question, we compared GRCNN with seven dynamic GNN models, including GCLSTM, AGCLSTM, DynamicTriad, STGCN, ASTGCN, DCRNN, and DGCNN, on the two datasets. All the eight models took as input node embeddings generated by JRWalk. Note that GRCNN-noatt denotes a GRCNN model built without the attention mechanism. Their prediction results in the DAP task are shown in Table \ref{table:node_GNN}. Three primary findings are described as follows.

First, because all the models in Table \ref{table:node_GNN} consider the temporal and spatial features simultaneously, they performed better than those node embedding approaches in Table \ref{table:node_embedding}. Second, AGCLSTM, ASTGCN, and GRCNN outperformed the corresponding models without attention, suggesting that the attention mechanism can effectively capture the changing features of input data. In particular, the comparison between GRCNN and GRCNN-noatt shows that the attention mechanism can capture the time dependencies among consecutive snapshots. Third, GRCNN achieved an accuracy of 75.5\% on the Eclipse dataset and an accuracy of 67.4\% on the Mozilla dataset. It was better than the seven baseline models in most cases. These observations indicate that GRCNN has the advantages over the SOTA models in extracting spatial-temporal features of dynamic networks.

\begin{table*}[]
\centering
\caption{Results of Different GNN Models (mean $\pm$ s. d.).}
\label{table:node_GNN}
\begin{tabular}{cllcclll}
\hline
\multirow{2}{*}{Model} & \multicolumn{3}{c}{Eclipse}                               & \multicolumn{3}{c}{Mozilla}                          & \multicolumn{1}{c}{}  \\ \cline{2-8}
                          & \multicolumn{1}{c}{40\%} & \multicolumn{1}{c}{60\%} & \multicolumn{1}{c}{80\%}     &  & \multicolumn{1}{c}{40\%} & \multicolumn{1}{c}{60\%} & \multicolumn{1}{c}{80\%} \\ \cline{1-8}
DynamicTriad              & 0.518$\pm$0.024 & 0.575$\pm$0.027 & 0.582$\pm$0.021 &  & 0.454$\pm$0.028 & 0.508$\pm$0.020 & 0.556$\pm$0.023 \\
GCLSTM                    & 0.536$\pm$0.033 & 0.583$\pm$0.031 & 0.602$\pm$0.019 &  & 0.461$\pm$0.032 & 0.531$\pm$0.025 & 0.563$\pm$0.026 \\
AGCLSTM                   & 0.553$\pm$0.023 & 0.591$\pm$0.025 & 0.613$\pm$0.018 &  & 0.482$\pm$0.028 & 0.540$\pm$0.024 & 0.581$\pm$0.021 \\
STGCN                     & 0.625$\pm$0.031 & 0.652$\pm$0.026 & 0.683$\pm$0.021 &  & 0.514$\pm$0.026 & 0.538$\pm$0.019 & 0.572$\pm$0.020 \\
ASTGCN                    & 0.650$\pm$0.027 & 0.676$\pm$0.023 & 0.692$\pm$0.017 &  & 0.550$\pm$0.026 & 0.566$\pm$0.020 & 0.613$\pm$0.018 \\
DCRNN                     & 0.664$\pm$0.035 & 0.680$\pm$0.026 & 0.718$\pm$0.024 &  & 0.573$\pm$0.043 & 0.611$\pm$0.032 & 0.643$\pm$0.025 \\
DGCNN                     & 0.684$\pm$0.032 & 0.703$\pm$0.027 & 0.732$\pm$0.021 &  & \textbf{0.609$\pm$0.028} & 0.621$\pm$0.019 & 0.649$\pm$0.021 \\
GRCNN-noatt               & 0.688$\pm$0.026 & 0.711$\pm$0.022 & 0.745$\pm$0.024 &  & 0.589$\pm$0.027 & 0.629$\pm$0.020 & 0.653$\pm$0.022 \\
GRCNN                     & \textbf{0.702$\pm$0.024} & \textbf{0.734$\pm$0.023} & \textbf{0.755$\pm$0.021} &  & 0.607$\pm$0.025 & \textbf{0.645$\pm$0.022} & \textbf{0.674$\pm$0.021} \\ \hline
\end{tabular}
\end{table*}

\subsubsection{Results of Different Developer Recommendation Approaches} \label{dev_rec}

For the third research question, we compared ST-DGNN with five baseline approaches (i.e., BOW+SVM, TGN, DF+SVM, DeepTriage, and ITriage) on the two datasets. BOW+SVM and DeepTriage utilize only bug reports’ text features, while the other four approaches leverage both text features from bug reports and structure features from DCNs. Their prediction results regarding accuracy@Top-$k$ in the BFP task are listed in Table \ref{table:node_recommendation}. Due to space limitations, we present only the Top-1, Top-3, and Top-5 recommendation results in Table \ref{table:node_recommendation}. 

It has been recognized that the comprehensive utilization of DCNs’ structure features and bug reports’ text features can be helpful to improve fixer recommendation accuracy. Because BOW+SVM and DeepTriage consider only the textual content of bug reports, their prediction accuracy values were lower than those of the other four methods on the two datasets. ST-DGNN achieved a 68.0\% (51.1\%) accuracy for the Top-1 recommendation, 77.4\% (66.2\%) accuracy for the Top-3 recommendation, and 87.5\% (75.4\%) accuracy for the Top-5 recommendation on the Eclipse (or Mozilla) dataset with an 80:20 split. It is obvious from Table \ref{table:node_recommendation} that ST-DGNN performs the best among the six recommendation approaches, followed by DF+SVM and TGN. In the same split setting, the Top-5 recommendation accuracy of DF+SVM achieved 83.6\% and 69.3\% on the Eclipse and Mozilla datasets, respectively, and TGN’s Top-5 recommendation accuracy was up to 79.3\% on the Eclipse dataset and 66.1\% on the Mozilla dataset. Across the two datasets, our approach, on average, improved the prediction accuracy of TGN by 10.7\% and DF+SVM by 7.9\%. These observations indicate that ST-DGNN can achieve the SOTA performance in recommending appropriate developers for new bug reports.

Although TGN, DF+SVM, and ITriage leverage text features from bug reports and structure features from DCNs, the former two approaches obtained better prediction results. The reasons are explained as follows. TGN uses a Markov chains model to reconstruct bug tossing graphs, which can reduce the redundant tossing paths between developers and optimize the network structure of a DCN. DF+SVM extracts high-level developer features from a few commonly-used aspects, which can better characterize the relationship between bug reports and developer attributes. ITriage uses a sequence-to-sequence model to learn interactive features from tossing sequences between developers while ignoring the complex topological structure of DCNs.

\begin{table*}[t]
\centering
\caption{Results of Different Developer Recommendation Approaches (mean $\pm$ s. d.).}
\label{table:node_recommendation}
\begin{tabular}{lllllclll}
\hline
                       & \multicolumn{1}{c}{\multirow{2}{*}{Approach}} & \multicolumn{3}{c}{Eclipse}                                                    & \multicolumn{3}{c}{Mozilla}                                                & \multicolumn{1}{c}{}     \\ \cline{3-9}
                       & \multicolumn{1}{c}{}                          & \multicolumn{1}{c}{40\%} & \multicolumn{1}{c}{60\%} & \multicolumn{1}{c}{80\%} &                      & \multicolumn{1}{c}{40\%} & \multicolumn{1}{c}{60\%} & \multicolumn{1}{c}{80\%} \\ \hline
\multirow{5}{*}{Top-1} & BOW+SVM                                       & 0.106$\pm$0.007 & 0.124$\pm$0.009 & 0.185$\pm$0.004 &  & 0.063$\pm$0.010 & 0.117$\pm$0.012 & 0.139$\pm$0.005 \\
                       & TGN                                           & 0.480$\pm$0.041 & 0.550$\pm$0.024 & 0.604$\pm$0.030 &  & 0.331$\pm$0.051 & 0.372$\pm$0.030 & 0.401$\pm$0.026 \\
                       & DF+SVM                                        & 0.513$\pm$0.012 & 0.592$\pm$0.011 & 0.632$\pm$0.009 &  & 0.376$\pm$0.023 & 0.412$\pm$0.012 & 0.464$\pm$0.003 \\
                       & DeepTriage                                    & 0.112$\pm$0.034 & 0.132$\pm$0.025 & 0.238$\pm$0.013 &  & 0.087$\pm$0.032 & 0.161$\pm$0.025 & 0.179$\pm$0.014 \\
                       & ITriage                                       & 0.264$\pm$0.024 & 0.327$\pm$0.018 & 0.407$\pm$0.013 &  & 0.226$\pm$0.022 & 0.349$\pm$0.021 & 0.436$\pm$0.012 \\ 
                       & ST-DGNN                                         & \textbf{0.542$\pm$0.023} & \textbf{0.615$\pm$0.016} & \textbf{0.680$\pm$0.017} &  & \textbf{0.416$\pm$0.026} & \textbf{0.470$\pm$0.021} & \textbf{0.511$\pm$0.014} \\\hline
\multirow{5}{*}{Top-3} & BOW+SVM                                       & 0.203$\pm$0.010 & 0.224$\pm$0.007 & 0.271$\pm$0.011 &  & 0.126$\pm$0.014 & 0.163$\pm$0.009 & 0.215$\pm$0.002 \\
                       & TGN                                           & 0.610$\pm$0.023 & 0.651$\pm$0.026 & 0.706$\pm$0.021 &  & 0.486$\pm$0.038 & 0.524$\pm$0.019 & 0.576$\pm$0.012 \\
                       & DF+SVM                                        & 0.621$\pm$0.015 & 0.690$\pm$0.005 & 0.722$\pm$0.007 &  & 0.503$\pm$0.018 & 0.545$\pm$0.014 & 0.582$\pm$0.004 \\
                       & DeepTriage                                    & 0.214$\pm$0.022 & 0.225$\pm$0.013 & 0.297$\pm$0.012 &  & 0.196$\pm$0.026 & 0.281$\pm$0.016 & 0.306$\pm$0.018 \\
                       & ITriage                                       & 0.380$\pm$0.022 & 0.523$\pm$0.018 & 0.611$\pm$0.016 &  & 0.387$\pm$0.024 & 0.572$\pm$0.013 & 0.608$\pm$0.014 \\ 
                       & ST-DGNN                                         & \textbf{0.649$\pm$0.012} & \textbf{0.713$\pm$0.015} & \textbf{0.774$\pm$0.018} &  & \textbf{0.563$\pm$0.022} & \textbf{0.613$\pm$0.017} & \textbf{0.662$\pm$0.018} \\\hline
\multirow{5}{*}{Top5}  & BOW+SVM                                       & 0.246$\pm$0.014 & 0.271$\pm$0.011 & 0.312$\pm$0.005 &  & 0.172$\pm$0.006 & 0.201$\pm$0.010 & 0.253$\pm$0.004 \\
                       & TGN                                           & 0.673$\pm$0.024 & 0.716$\pm$0.027 & 0.793$\pm$0.019 &  & 0.553$\pm$0.026 & 0.604$\pm$0.023 & 0.661$\pm$0.016 \\
                       & DF+SVM                                        & 0.712$\pm$0.017 & 0.754$\pm$0.012 & 0.836$\pm$0.005 &  & 0.586$\pm$0.014 & 0.632$\pm$0.007 & 0.693$\pm$0.011 \\
                       & DeepTriage                                    & 0.269$\pm$0.023 & 0.283$\pm$0.021 & 0.342$\pm$0.014 &  & 0.261$\pm$0.023 & 0.305$\pm$0.014 & 0.327$\pm$0.013 \\
                       & ITriage                                       & 0.481$\pm$0.025 & 0.604$\pm$0.017 & 0.642$\pm$0.014 &  & 0.474$\pm$0.018 & 0.554$\pm$0.021 & 0.634$\pm$0.015 \\ 
                       & ST-DGNN                                         & \textbf{0.730$\pm$0.019} & \textbf{0.802$\pm$0.013} & \textbf{0.875$\pm$0.012} &  & \textbf{0.657$\pm$0.020} & \textbf{0.711$\pm$0.016} & \textbf{0.754$\pm$0.017} \\\hline
\end{tabular}
\end{table*}

\section{Discussion} \label{sec:Discussion}

\subsection{Comparison of GRCNN Variants}

We further analyzed the impacts of GRCNN’s three components on prediction performance in the DAP and BFP tasks. We trained three new GRCNN models with a single component while deactivating the other two components. GRCNN-hour, GRCNN-day, and GRCNN-week denote a GRCNN model that considers only the hourly-periodic component, daily-periodic component, and weekly-periodic component, respectively. Also, we repeated the Monte Carlo Cross-Validation approach ten times and calculated the average performance in terms of accuracy. The prediction results of GRCNN-hour, GRCNN-day, and GRCNN-week are shown in Table \ref{table:node_time}. 

GRCNN-day achieved a 71.3\% accuracy and 65.2\% accuracy on the Eclipse dataset and Mozilla dataset, respectively, with an 80:20 split in the DAP task. Its prediction accuracy values were, on average, higher than those of GRCNN-hour and GRCNN-week. This result suggests that the daily-periodic component contributes more to prediction performance. In the BFP task, GRCNN-week performed better than the other two models, and its accuracy was 3.2\% and 5.1\% higher than that of GRCNN-day and GRCNN-hour, respectively. This result suggests that the weekly-periodic component is more useful in developer recommendation.

\begin{table*}[t]
\centering
\caption{Comparison of GRCNN variants (mean $\pm$ s. d.).}
\label{table:node_time}
\begin{tabular}{lclllclll}
\hline
\multirow{2}{*}{Task} & \multirow{2}{*}{Model} & \multicolumn{3}{c}{Eclipse}                                                    & \multicolumn{3}{c}{Mozilla}                                                & \multicolumn{1}{c}{}     \\ \cline{3-9} 
                      &                           & \multicolumn{1}{c}{40\%} & \multicolumn{1}{c}{60\%} & \multicolumn{1}{c}{80\%} &                      & \multicolumn{1}{c}{40\%} & \multicolumn{1}{c}{60\%} & \multicolumn{1}{c}{80\%} \\ \hline
\multirow{3}{*}{DAP}  & GRCNN-hour                & 0.597$\pm$0.033 & 0.613$\pm$0.026 & 0.641$\pm$0.025 &  & 0.548$\pm$0.023 & 0.593$\pm$0.032 & 0.621$\pm$0.026 \\
                      & GRCNN-day                 & \textbf{0.632$\pm$0.028} & 0.671$\pm$0.031 & \textbf{0.713$\pm$0.024} &  & 0.576$\pm$0.024 & \textbf{0.639$\pm$0.027} & \textbf{0.652$\pm$0.025} \\
                      & GRCNN-week                & 0.614$\pm$0.027 & \textbf{0.676$\pm$0.024} & 0.696$\pm$0.026 &  & \textbf{0.582$\pm$0.026} & 0.611$\pm$0.030 & 0.641$\pm$0.028 \\ \hline
\multirow{3}{*}{BFP}  & GRCNN-hour                & 0.498$\pm$0.045 & 0.574$\pm$0.028 & 0.645$\pm$0.023 &  & 0.372$\pm$0.038 & 0.426$\pm$0.042 & 0.463$\pm$0.033 \\
                      & GRCNN-day                 & 0.519$\pm$0.037 & 0.593$\pm$0.021 & 0.661$\pm$0.026 &  & 0.391$\pm$0.028 & 0.443$\pm$0.027 & \textbf{0.495$\pm$0.032} \\
                      & GRCNN-week                & \textbf{0.531$\pm$0.026} & \textbf{0.619$\pm$0.022} & \textbf{0.664$\pm$0.021} &  & \textbf{0.423$\pm$0.030} & \textbf{0.451$\pm$0.029} & 0.487$\pm$0.027  \\ \hline
\end{tabular}
\end{table*}

\subsection{Parameter Sensitivity Analysis} 

\begin{figure*}[hbt!]
\centering
\includegraphics[width=4.5in]{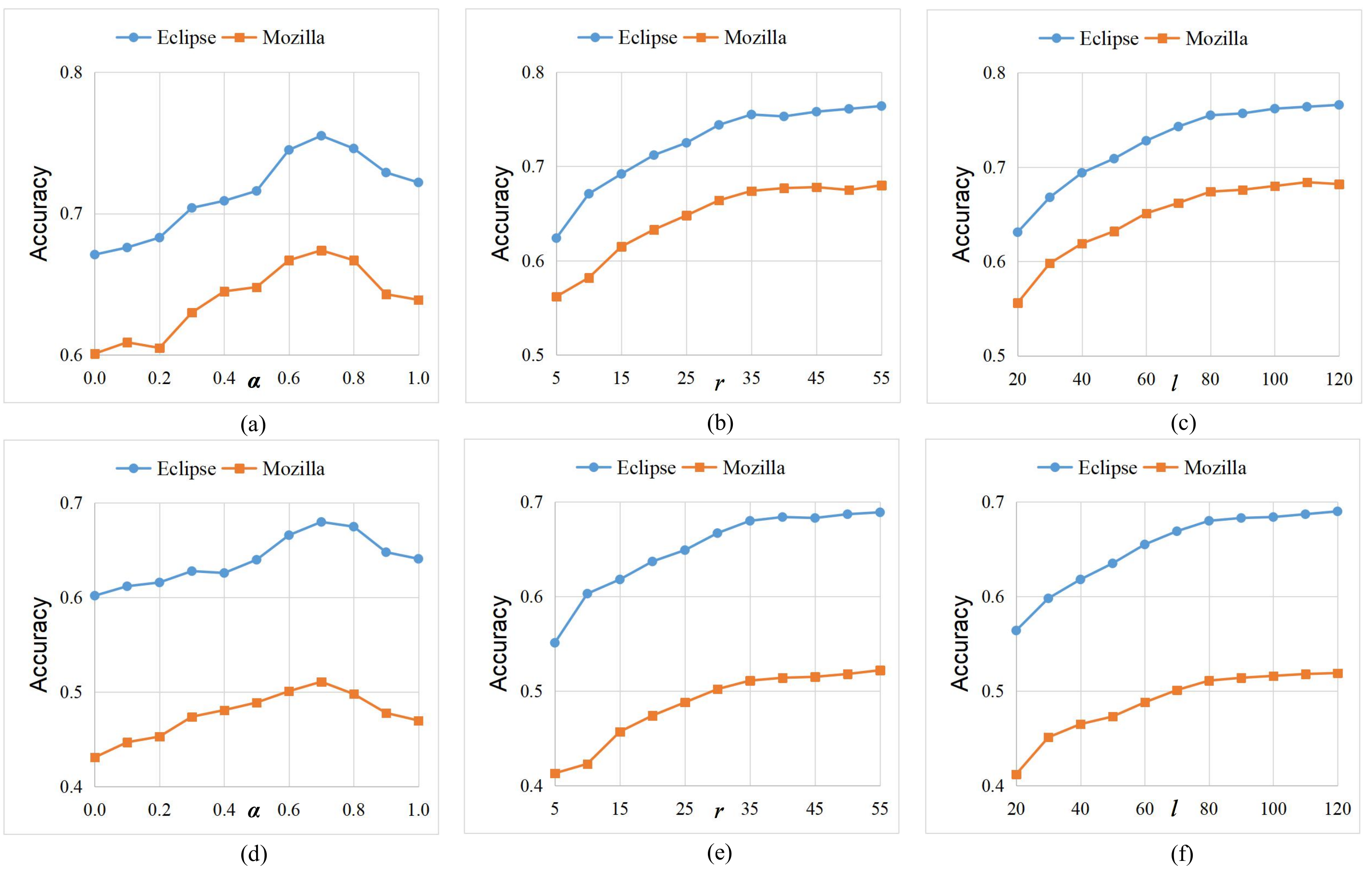}
\caption{Parameter sensitivity study on the probability $\alpha$, the number of walks per node $r$, and the length of JRWalk $l$. Plots (a)-(c) present JRWalk’s results in the DAP task, and plots (d)-(f) show ST-DGNN’s results in the BFP task.}
\label{param_sentivity}
\end{figure*}

We also analyzed the sensitivity of parameters $\alpha$, $r$, and $l$ in the DAP and BFP tasks. The first parameter $\alpha$ denotes the probability of performing joint random walks on a graph, and the higher value of $\alpha$ means the preference of edge-importance-based random walks. In Fig. \ref{param_sentivity}(a) and Fig. \ref{param_sentivity}(d), we plot the two tasks’ prediction results by JRWalk and ST-DGNN as a function of $\alpha$ on the Eclipse and Mozilla datasets. If $\alpha$ = 0, JRWalk is purely based on node importance. The prediction accuracy and Top-1 fixer recommendation accuracy increase with the increase of $\alpha$. They reach the maximum when $\alpha$ is equal to 0.7 and then decrease with the increase of $\alpha$. JRWalk degenerates into an edge-importance-based random walk mechanism when $\alpha$ = 1.

The second parameter $r$ denotes the number of walking sequences that we sample for each node. Fig. \ref{param_sentivity}(b) and Fig. \ref{param_sentivity}(e) show that $r$ has a positive correlation with the two tasks’ prediction results by JRWalk and ST-DGNN. We changed the value of $r$ from five to fifty-five and found that the two tasks’ prediction performance remained relatively stable after $r > 35$. Considering that a higher value of $r$ requires more computing resources, $r$ was set to 35 in the experiments after making a trade-off between performance and computation cost.

The third parameter $l$ represents the length of a walking sequence, i.e., the number of nodes in a walking sequence. Larger $l$ means that a node has more opportunities to interact with distant nodes. In the experiments, we varied the value of $l$ from 20 to 120. As shown in Fig. \ref{param_sentivity}(c) and Fig. \ref{param_sentivity}(f), the prediction accuracy of the two tasks also has a positive correlation with this parameter, but the growth rate becomes very small after $l$ exceeds 90. Due to the trade-off between performance and computation cost, $l$ was set to 90 in the experiments.

\section{Threats to Validity} \label{sec:Threats}

In this section, we discuss some potential threats to the validity of the experimental results in this article.

{\it Internal Validity}: There are two main threats to the internal validity of our work: evaluation
criterion and parameter settings. First, there is no recognized evaluation criterion for graph representation learning. Previous studies usually fed the latent representations generated by different approaches to specific tasks, such as node classification and link prediction, to test their effectiveness. Therefore, in this study, we compared our approach and selected baseline methods in two domain-specific tasks that belong to node classification. Second, the parameters of all baseline approaches were set following their original studies or set to the default parameter settings, without additional optimization. We also believe that the experimental results may change by optimizing or fine-tuning these baselines’ parameters.

{\it External Validity}: The external validity concerns the generalizability of our approach. First, some new approaches proposed in 2020, especially without implementation code, were not involved in the comparison experiments. Second, the experimental results conducted on the Eclipse and Mozilla projects indicate the advantages of our approach. However, the bug-triaging performance of the ST-DGNN framework remains unknown for other large-scale open-source and closed-source software projects. Third, the JRWalk mechanism aims to embed nodes in homogeneous networks. Therefore, it cannot directly apply to heterogeneous networks, which have recently attracted more attention. Our future work is to improve further the proposed framework by extending JRWalk for dynamic heterogeneous networks.

\section{Conclusion} \label{sec:Conclusion}

In this article, we propose a novel spatial-temporal dynamic graph neural network framework for two domain-specific tasks on developer collaboration networks. To improve random-walk-based graph representation learning approaches, we put forward a more informative walking mechanism, joint random walk, which can perform more effective sampling by considering both node importance and edge importance. Inspired by the recent studies on dynamic graph neural networks, we further design a graph recurrent convolutional neural network model based on joint random walks and an attention mechanism to conduct graph representation on dynamic networks. The proposed model also applies a periodic sampling trick to optimize a dynamic network’s temporal features at different time slices. Experiments on two large-scale open-source software projects indicate that our approach is better than all the baseline approaches in the two tasks.

Since multiple types of interactions (such as email communication and social activity) exist among developers, multiplex DCNs can be constructed for further investigation. We plan to extend JRWalk and analyze its embedding performance on multiplex networks or heterogeneous networks. The multiple interactions of developers change over time. Another practical work would be exploring a spatial-temporal feature extractor (or graph representation model) on multiplex dynamic DCNs.

% or


%\appendices

%\section{}

\section*{Acknowledgment}

This work was supported by the National Key Research and Development Program of China (No. 2018YFB1003801), the National Science Foundation of China (No. 61972292), and the Natural Science Foundation of Fujian Province in China (No. 2020J01816). Yutao Ma is the corresponding author of this paper.

\ifCLASSOPTIONcaptionsoff
  \newpage
\fi



%\bibliographystyle{IEEEtran}

\bibliographystyle{IEEEtran}
\bibliography{ref}

%\begin{thebibliography}{1}
%\end{thebibliography}


%\begin{IEEEbiography}{Hongrun Wu}
%\end{IEEEbiography}


%\newpage



%\vfill


\end{document}